\documentclass[a4paper,final,confe rence]{IEEEtran}
\usepackage{mathpple}
\usepackage{times}
\usepackage{capt-of}
\usepackage[top=0.55in, bottom=0.55in, left=0.55in, right=0.55in]{geometry}

\usepackage{amsmath}  
\usepackage{amssymb}  
\usepackage{mathrsfs} 
\usepackage{kantlipsum,cuted}
\usepackage{multirow}
\usepackage{tabu}

\usepackage{lipsum}

\usepackage{theorem}  
\usepackage{cite}     
\usepackage{comment}  

\usepackage{upref}
\usepackage{amsfonts}

\usepackage{verbatim}

\usepackage[dvipsnames,usenames]{color}
\usepackage{enumerate}
\usepackage{multicol}
\usepackage{blkarray, bigstrut}
\usepackage{graphicx}
\usepackage{subfigure}

\usepackage{latexsym}

\usepackage[draft,ulem=normalem]{changes}
\usepackage[all]{xy}

\usepackage{algorithmicx}
\usepackage{algorithm}
\usepackage[noend]{algpseudocode}
\algrenewcommand\alglinenumber[1]{\scriptsize #1:}
\algrenewcommand\algorithmicindent{1em}%

\usepackage{wrapfig}
\usepackage[shortlabels]{enumitem}

\makeatletter
\def\BState{\State\hskip-\ALG@thistlm}

\makeatother

\newcommand{\be}[1]{\begin{equation}\label{#1}}
	\newcommand{\ee}{\end{equation}}

\newcommand{\bc}{\begin{center}}
	\newcommand{\ec}{\end{center}}

\newcommand{\ceil}[1]{\lceil{#1}\rceil}

\newcommand{\cC}{{\cal C}}
\newcommand{\cD}{{\cal D}}

\newcommand{\cL}{{\cal L}}

\newcommand{\cS}{{\cal S}}

\newcommand{\bfc}{{\boldsymbol c}}

\newcommand{\bfx}{{\boldsymbol x}}
\newcommand{\bfy}{{\boldsymbol y}}

\DeclareMathOperator*{\argmax}{arg\,max}

\renewcommand{\le}{\leqslant}
\renewcommand{\leq}{\leqslant}
\renewcommand{\ge}{\geqslant}
\renewcommand{\geq}{\geqslant}

\newcommand{\Cref}[1]{Co\-rol\-la\-ry\,\ref{#1}}

\theoremstyle{plain} \theorembodyfont{\normalfont\slshape}

\newtheorem{thm}{Theorem$\!$}
\newenvironment{theorem}{\begin{thm}\hspace*{-1ex}{\bf.}}{\end{thm}}

\newtheorem{prop}[thm]{Proposition$\!$}

\newtheorem{lem}[thm]{Lemma$\!$}
\newenvironment{lemma}{\begin{lem}\hspace*{-1ex}{\bf.}}{\end{lem}}

\newtheorem{cor}[thm]{Corollary$\!$}
\newenvironment{corollary}{\begin{cor}\hspace*{-1ex}{\bf.}}{\end{cor}}

\newtheorem{prob}[thm]{Problem$\!$}

\newtheorem{cl}[thm]{Claim$\!$}
\newenvironment{claim}{\begin{cl}\hspace*{-1ex}{\bf.}}{\end{cl}}

\newtheorem{defi}[thm]{Definition$\!$}

\theorembodyfont{\normalfont}

\newtheorem{exam}{Example$\!$}

\newtheorem{remrk}{Remark$\!$}

\definecolor{Codecolor}{named}{White}  

	\newcommand{\Copen}{\mbox{\{\kern-5.50pt\{}}
	\newcommand{\Cclose}{\mbox{\}\kern-5.50pt\}}}
	\newcommand{\Cslash}{\mbox{$\backslash\kern-6.02pt\backslash$}}

\newcommand{\RN}[1]{%
	\textup{\uppercase\expandafter{\romannumeral#1}}%
}

\newtheorem{innercustomgeneric}{\customgenericname}
\providecommand{\customgenericname}{}
\newcommand{\newcustomtheorem}[2]{%
	\newenvironment{#1}[1]
	{%
		\renewcommand\customgenericname{#2}%
		\renewcommand\theinnercustomgeneric{##1}%
		\innercustomgeneric
	}
	{\endinnercustomgeneric}
}
\newcustomtheorem{customproposition}{Proposition}
\newcustomtheorem{customtheorem}{Theorem}
\newcustomtheorem{customlemma}{Lemma}
\newcustomtheorem{customclaim}{Claim}

\newcommand{\pr}{\ensuremath{\mathsf{Pr}}}
\newcommand{\ch}{\ensuremath{\mathsf{S}}}
\newcommand{\bsc}{\ensuremath{\mathsf{BSC}}}

\newcommand{\zc}{\ensuremath{\mathsf{Z}}}
\newcommand{\del}{\ensuremath{\mathsf{Del}}}
\newcommand{\ins}{\ensuremath{\mathsf{Ins}}}
\newcommand{\emb}{\ensuremath{\mathsf{Emb}}}
\newcommand{\perr}{\ensuremath{\mathsf{P_{err}}}}
\newcommand{\pfail}{\ensuremath{\mathsf{P_{fail}}}}
\newcommand{\prun}{\ensuremath{\mathsf{P_{run}}}}
\newcommand{\palt}{\ensuremath{\mathsf{P_{alt}}}}
\newcommand{\cp}{\ensuremath{\mathsf{Cap}}}
\newcommand{\ent}{\ensuremath{\mathsf{H}}}
\newcommand{\sups}{\ensuremath{\cS\mathsf{CS}}}
\newcommand{\subs}{\ensuremath{\cL\mathsf{CS}}}
\newcommand{\ML}{\ensuremath{\mathsf{ML}}}
\newcommand{\perrin}{\ensuremath{\mathsf{P_{err}^{ins}}}}
\newcommand{\pfailin}{\ensuremath{\mathsf{P_{fail}^{ins}}}}	
\newcommand{\prunin}{\ensuremath{\mathsf{P_{run}^{ins}}}}
\newcommand{\paltin}{\ensuremath{\mathsf{P_{alt}^{ins}}}}
	
\newcommand{\e}[1]{\textcolor{red}{#1}}

\begin{document}
		\title{\textbf{The Error Probability of Maximum-Likelihood Decoding over Two Deletion Channels}\vspace{-0.2ex}}
		\author{\IEEEauthorblockN{\textbf{Omer Sabary}}
			\IEEEauthorblockA{Dept. of Computer Science\\
				\hspace{-3ex}Technion --- Israel Inst. of Technology\\
				Haifa 3200003, Israel \\
				omersabary@cs.technion.ac.il\vspace{-4.2ex}}
			\and 
			\IEEEauthorblockN{\textbf{Eitan Yaakobi}}
		\IEEEauthorblockA{Dept. of Computer Science\\
				\hspace{0ex}Technion --- Israel Inst. of Technology\\
				Haifa 3200003, Israel \\
				yaakobi@cs.technion.ac.il\vspace{-4.2ex}}
			\and
			\IEEEauthorblockN{\textbf{Alexander Yucovich}}
			\IEEEauthorblockA{Dept. of Computer Science\\
				Technion --- Israel Inst. of Technology\\
				Haifa 3200003, Israel \\
				yucovich@gmail.com\vspace{-4.2ex}}}
		\maketitle
\begin{abstract}
This paper studies the problem of reconstructing a word given several of its noisy copies. This setup is motivated by several applications, among them is reconstructing strands in DNA-based storage systems. Under this paradigm, a word is transmitted over some fixed number of identical independent channels and the goal of the decoder is to output the transmitted word or some close approximation. The main focus of this paper is the case of two deletion channels and studying the error probability of the maximum-likelihood (ML) decoder under this setup. First, it is discussed how the ML decoder operates. Then, we observe that the dominant error patterns are deletions in the same run or errors resulting from alternating sequences. Based on these observations, it is derived that the error probability of the ML decoder is roughly $\frac{3q-1}{q-1}p^2$, when the transmitted word is any $q$-ary sequence and $p$ is the channel's deletion probability. We also study the cases when the transmitted word belongs to the Varshamov Tenengolts (VT) code or the shifted VT code. Lastly, the insertion channel is studied as well. These theoretical results are verified by corresponding simulations. 
\end{abstract}
\renewcommand{\baselinestretch}{0.95}\normalsize
\section{Introduction} \label{sec:intro}

\emph{Reconstruction of sequences} refers to a large class of problems in which there are several noisy copies of the information and the goal is to decode the information, either with high probability or in the worst case. One of the more relevant models to the study in this paper is the \emph{trace reconstruction problem}~\cite{BKK04, PZ17, NP17, HMP08, HPP18}, where it is assumed that a sequence is transmitted through multiple deletion channels, so each bit is deleted with some fixed probability $p$. Under this setup, the goal is to determine the minimum number of traces, i.e., channels, required to reconstruct the sequence with high probability. Other examples include the \emph{sequence reconstruction problem} which was first studied by Levenshtein~\cite{L011,L012,gabrys2018sequence, yaakobi2013sequence, yaakobi2012uncertainty, sala2015three, levenshtein2009error,  levenshtein2008reconstruction}. One of the dominant motivating applications of the sequence reconstruction problems is DNA storage~\cite{CGK12, OAC17, YGM16, GHP15, Anavy433524}, where every DNA strand has several noisy copies. 

Many of the reconstruction problems are focused on studying the minimum number of channels required for successful decoding. However, in many cases, the number of channels is fixed and then the goal is to find the best code construction that is suitable for this channels setup. Motivated by this important observation, the goal of this paper is to study the error probability of maximum-likelihood decoding when a word is transmitted over two deletion or insertion channels. This study is also motivated by the recent works of Srinivasavaradhan et. al.~\cite{srinivasavaradhan2018maximum,srinivasavaradhan2019symbolwise}, where reconstruction algorithms for the maximum-likelihood have been studied. Abroshan et al. presented in~\cite{AVDG19} a new coding scheme for sequence reconstruction which is based on the Varshamov Tenegolts (VT) code~\cite{VT} and in a parallel work~\cite{ISIT20} it is studied how to design codes for the worst case, when the number of channels is given. 

When a word is transmitted over the deletion channel,  the channel output is necessary a subsequence of the transmitted word. Hence, when transmitting the same word over multiple deletion channels, the possible candidate words for decoding are the so-called \emph{common supersequences} of all channels' outputs. Hence, an important part of the decoding process is to find the set of all possible common supersequences and in particular the \emph{shortest common supersequences} (\emph{SCS})~\cite{itoga1981string}. Even though this problem is in general NP hard~\cite{B93} for an arbitrary number of sequences, for two words a dynamic programming algorithm exists with quadratic complexity; see~\cite{itoga1981string} for more details and further improvements and approximations for two or more sequences~\cite{Irving1994,hirschberg1975linear,tronicek1999problems, ukkonen1990linear}. The case of finding the \emph{longest common subsequences} (\emph{LCS}) is no less interesting and has been extensively studied in several previous works; see e.g.~\cite{hirschberg1977algorithms, apostolico1992fast, hsu1984computing, masek1980faster, sankoff1972matching, chen2006fast}. Most of these works focused on improving the complexity of the dynamic programming algorithm suggested in~\cite{apostolico1992fast} and presented heuristics and approximations for the LCS.

When a sequence is transmitted over two (or more) deletion channels, the first step in the ML decoding algorithm is to build upon the algorithm for finding the SCS to generate all possible candidate words, i.e., all shortest common supersequences. However, if there is more than one candidate, it is necessary to find the one that maximizes the probability that it was sent over all channels. It is shown that this problem is directly related to finding the \emph{embedding number} between two sequences~\cite{DBLP:journals/corr/abs-1802-00703,elzinga2008algorithms}. This value is the number of times a sequence can be generated by deletions from its supersequence. 

These two steps are in fact the process in which the ML decoder operates. Thus, the main goal of this paper is to determine the error probability of the ML decoder. Assume the deletion probability of every channel is $p$. If there are $t$ channels, then a lower bound on the error probability is at least $p^t$ since if a bit is deleted in all $t$ channels, then it will also be deleted at the ML decoder's output. However, it will be observed that this lower bound is not tight. For example, if there are two channels and a bit is deleted in a run in both channels (i.e., not necessarily the same bit), then this run will have a deletion error as well. This indeed will be one of the main error patterns of the ML decoder. Furthermore, it will be observed that alternating subsequences in the word are also error prone and these two will be the only dominant error patterns. Thus, we will show that for arbitrary $q$-ary sequences, the error decoding probability for the runs is $\frac{q+1}{q-1}p^2$, while for the alternating sequences is $2p^2$, independently of the field size.

The rest of the paper is organized as follows. Section~\ref{sec:defs} presents the notations and the formal definition of transmission over multiple channels which will be studied in the paper. Section~\ref{sec:del_ch} studies this problem for the insertion and deletion channels. In Section~\ref{sec:two deletions}, we present our main results for the case of two channels. We consider the average decoding failure probability of the ML decoder and its average decoding error probability when the code is the entire space, the VT code, and the shifted VT code. We then continue in Section~\ref{sec:ins} to study the equivalent problem for insertions. Section~\ref{sec:conc} concludes the paper and discusses open problems. Due to the lack of space, some of the proofs in the paper are omitted.

\section{Definitions and Preliminaries}\label{sec:defs}		

We denote by $\Sigma_q =\{0,\ldots,q-1\}$ the alphabet of size $q$ and $\Sigma_q^* \triangleq \bigcup_{\ell=0}^\infty \Sigma_q^\ell, \Sigma_q^{\leq n} \triangleq \bigcup_{\ell=0}^n \Sigma_q^\ell, \Sigma_q^{\geq n} \triangleq \bigcup_{\ell=n}^\infty \Sigma_q^\ell$. 
The length of $\bfx\in \Sigma^n$ is denoted by $|\bfx|=n$. 
The \emph{Levenshtein distance}  between two words $\bfx,\bfy \in \Sigma_q^*$, denoted by $d_L(\bfx,\bfy)$, is the minimum number of insertions and deletions required to transform $\bfx$ into $\bfy$, and
$d_H(\bfx,\bfy)$ denotes the \emph{Hamming distance} between $\bfx$ and $\bfy$, when $|\bfx| = |\bfy|$. 
A word $\bfx\in\Sigma_q^*$ will be referred as an \emph{alternating sequence} if it cyclically repeats all symbols in $\Sigma_q$ in the same order. For example, for $\Sigma_2=\{0,1\}$, the two alternating sequences are $010101\cdots$ and $101010\cdots$, and in general there are $q!$ alternating sequences.  For $n\geq 1$, the set $\{1,\ldots,n\}$ is abbreviated by $[n]$. 

For a word $\bfx \in \Sigma_q^*$ and a set of indices $I\subseteq [|\bfx|]$, the word $\bfx_I$ is the \emph{projection} of $\bfx$ on the indices of $I$ which is the subsequence of $\bfx$ received by the symbols in the entries of $I$. A word $\bfx \in \Sigma^*$ is called a \emph{supersequence} of $\bfy \in \Sigma^*$, if $\bfy$ can be obtained by deleting symbols from $\bfx$, that is, there exists a set of indices $I\subseteq [|\bfx|]$ such that $\bfy = \bfx_I$. In this case, it is also said that $\bfy$ is a \emph{subsequence} of $\bfx$. Furthermore, $\bfx$ is called a \emph{common supersequence}, \emph{subsequence} of some words $\bfy_1,\ldots,\bfy_t$ if $\bfx$ is a supersequence, subsequence of each one of these $t$ words. 
The set of all common supersequences of $\bfy_1,\ldots,\bfy_t\in \Sigma_q^*$ is denoted by $\sups(\bfy_1,\ldots,\bfy_t)$ and $\mathsf{SCS}(\bfy_1,\dots,\bfy_t)$ is the \emph{length of the shortest common supersequence} (\emph{SCS)} of $\bfy_1,\dots,\bfy_t$, that is, 
$\mathsf{SCS}(\bfy_1,\dots,\bfy_t) = \min_{\bfx\in \sups(\bfy_1,\ldots,\bfy_t)}\{|\bfx|\}$. Similarly, $\subs(\bfy_1,\ldots,\bfy_t)$ is the set of all subsequences of $\bfy_1,\dots,\bfy_t$ and $\mathsf{LCS}(\bfy_1,\dots,\bfy_t)$ is the \emph{length of the longest common subsequence} (\emph{LCS)} of $\bfy_1,\dots,\bfy_t$, that is, $\mathsf{LCS}(\bfy_1,\dots,\bfy_t) = \max_{\bfx\in \subs(\bfy_1,\ldots,\bfy_t)}\{|\bfx|\}$.

We consider a channel $\ch$ that is characterized by a conditional probability $\pr_\ch$, which is defined by 
$$\pr_\ch\{  \bfy\textmd{ rec. }| \bfx \textmd{ trans.}\},$$
for every pair $(\bfx,\bfy)\in(\Sigma_q^*)^2$. 
Note that it is not assumed that the lengths of the input and output words are the same as we consider also deletions and insertions of symbols, which is the main topic of this work. As an example, it is well known that if $\ch$ is the \emph{binary symmetric channel} (\emph{BSC}) with crossover probability $0\leq p \leq 1/2$, denoted by $\bsc(p)$, it holds that 
$$\pr_{\bsc(p)}\{  \bfy\textmd{ rec. }| \bfx \textmd{ trans.}\} = p^{d_H(\bfy,\bfx)}(1-p)^{n-d_H(\bfy,\bfx)},$$
for all $(\bfx,\bfy)\in (\Sigma_2^n)^2$, and otherwise (the lengths of $\bfx$ and $\bfy$ is not the same) this probability equals 0. Similarly, for the $Z$-channel, denoted by $\zc(p)$, it is assumed that only a 0 can change to a 1 with probability $p$ and so 
$$\pr_{\zc(p)}\{  \bfy\textmd{ rec. }| \bfx \textmd{ trans.}\} = p^{d_H(\bfy,\bfx)}(1-p)^{n-d_H(\bfy,\bfx)},$$
for all $(\bfx,\bfy)\in (\Sigma_2^n)^2$ such that $\bfx \leq \bfy$, and otherwise this probability equals 0. 

In the \emph{deletion channel} with deletion probability $p$, denoted by $\del(p)$, every symbol of the word $\bfx$ is deleted with probability $p$. Similarly, in the \emph{insertion channel} with insertion probability $p$, denoted by $\ins(p)$, a symbol is inserted in each of the possible $|\bfx|+1$ positions of the word $\bfx$ with probability $p$, while the probability to insert each of the symbols in $\Sigma_q$ is the same and equals $\frac{p}{q}$. 

A decoder for a code $\cC$ with respect to the channel $\ch$ is a function $\cD:\Sigma_q^*\rightarrow \cC$. Its \emph{average decoding failure probability} is defined by $\pfail(\ch,\cC,\cD) = \frac{\sum_{\bfc\in\cC}\pfail(\bfc)}{|\cC|}$, where 
$$\pfail(\bfc) =   \sum_{\bfy:\cD(\bfy) \neq  \bfc} \pr_\ch\{  \bfy\textmd{ rec. }| \bfc \textmd{ trans.}\}.$$
We will also be interested in the \emph{average decoding error probability} which is the average normalized distance between the transmitted word and the decoder's output. The distance will depend upon the channel of interest. For example, for the BSC we will consider the Hamming distance, while for the deletion and insertion channels, the Levenshtein distance will be of interest. Hence, for a channel $\ch$, distance function $d$, and a decoder $\cD$, we let $\perr(\ch,\cC,\cD,d) = \frac{\sum_{\bfc\in\cC}\perr(\bfc,d)}{|\cC|}$, where 
$$\perr(\bfc,d) =   \sum_{\bfy:\cD(\bfy) \neq  \bfc} \frac{d(\bfy,\bfc)}{|\bfc|}\cdot \pr_\ch\{  \bfy\textmd{ rec. }| \bfc \textmd{ trans.}\}.$$

The \emph{maximum-likelihood} (\emph{ML}) \emph{decoder} for $\cC$ with respect to $\ch$, denoted by $\cD_{\ML}$, outputs a codeword $\bfc\in\cC$ that maximizes the probability $\pr_\ch\{  \bfy\textmd{ rec. }| \bfc \textmd{ trans.}\}$. That is, for  $\bfy\in \Sigma_{q}^*$, 
$$\cD_{\ML}(\bfy) =\argmax_{\bfc\in\cC}\left\{\{\pr_\ch\{  \bfy\textmd{ rec. }| \bfc \textmd{ trans.}\}\right\}.$$
It is well known that for the BSC, the ML decoder simply chooses the closet codeword with respect to the Hamming distance. 
The \emph{channel capacity} is referred as the maximum information rate that can be reliably transmitted over the channel $\ch$ and is denoted by $\cp(\ch)$. For example, 
$\cp(\bsc(p))=1-\ent(p)$, where $\ent(p)=-p\log(p)-(1-p)\log(1-p)$ is the binary entropy function. 

The conventional setup of channel transmission is extended to the case of more than a single instance of the channel. Assume a word $\bfx$ is transmitted over some $t$ identical channels of $\ch$ and the decoder receives all channel outputs $\bfy_1,\ldots,\bfy_t$. This setup is characterized by the conditional probability 
$$\pr_{(\ch,t)}\{ \bfy_1,\ldots,\bfy_t \textmd{ rec.}| \bfx \textmd{ trans.}\} = \prod_{i=1}^t\pr_{\ch}\{ \bfy_i\textmd{ rec.}| \bfx \textmd{ trans.}\}.$$
The definitions of a decoder, the ML decoder and the error probabilities are extended similarly. The input to the ML decoder is the words $\bfy_1,\ldots,\bfy_t$ and the output is the codeword $\bfc$ which maximizes the probability $\pr_{(\ch,t)}\{ \bfy_1,\ldots,\bfy_t \textmd{ rec.}| \bfx \textmd{ trans.}\}$. The average decoding failure probability, average decoding error probability is generalized in the same way and is denoted by  $\pfail(\ch,t,\cC,\cD)$, $\perr(\ch,t,\cC,\cD,d)$, respectively. The capacity of this channel is denoted by $\cp(\ch,t)$, so $\cp(\ch,1) = \cp(\ch)$.

The case of the BSC was studied by Mitzenmacher in~\cite{M06}, where he showed that 
\begin{small}
\begin{align*}
& \cp(\bsc(p),t))  & \\
& =1 +\sum_{i=0}^t\binom{t}{i} \left( p^i(1-p)^{t-i}\log \frac{p^i(1-p)^{d-i}}{p^i(1-p)^{t-i}+p^{t-i}(1-p)^i} \right).&
\end{align*}\vspace{-2ex}
\end{small}

\noindent On the other hand, the $Z$ channel is significantly easier to solve and it is possible to verify that  $\cp(\zc(p),t) = \cp(\zc(p^t))$.
It is also possible to calculate the average decoding error and failure probabilities for the BSC and $Z$ channels. For example, when $\cC=\Sigma_2^n$, one can verify that 
$$\perr(\zc(p),t,\Sigma_2^n,\cD_{\ML},d_H) = p^t,$$
and if $t$ is odd then
$$\perr(\bsc(p),t,\Sigma_2^n,\cD_{\ML},d_H) = \sum_{i=0}^{\frac{t-1}{2}} \binom{t}{i} p^{t-i}(1-p)^i.$$
Similarly, $\pfail(\zc(p),t,\Sigma_2^n,\cD_{\ML}) = 1- (1-p^t)^n$ and $\pfail(\bsc(p),t,\Sigma_2^n,\cD_{\ML}) \hspace{-0.25ex}=1-\hspace{-0.25ex} (1 \hspace{-0.25ex}-\hspace{-0.25ex}   \sum_{i=0}^{\frac{t-1}{2}} \binom{t}{i} p^{t-i}(1\hspace{-0.25ex}-\hspace{-0.25ex}p)^i )^n$.
However, calculating these probabilities for the deletion and insertion channels is a far more challenging task. The goal of this paper is to
study in depth the special case of $t=2$ and estimate the average error and failure probabilities, when the code is the entire space, the Varshamov Tenengolts (VT) code~\cite{VT}, and the shifted VT (SVT) code~\cite{7837631}. 

This model is closely connected to several related problems. In the \emph{reconstruction problem} studied by Levenshtein~\cite{L011,L012}, it was assumed that the word is transmitted over several noisy channels and the goal of the decoder is to decode the transmitted word in the worst case, assuming that all channels' outputs are different from each other. Several extensions of these problems have been studied; see e.g.~\cite{gabrys2018sequence, yaakobi2013sequence, yaakobi2012uncertainty, sala2015three, levenshtein2009error,  levenshtein2008reconstruction}, 
however in all of them the goal is to find the number of channel that guarantees unique decoding in the worst case. The most relevant case of the reconstruction problem to our work is the one studied in~\cite{8294216}, where it was shown how the shifted VT codes can be used for the two single-deletion channels case. In our parallel work~\cite{ISIT20} the dual problem is studied where the number of channels is given and then the goal is to find the best code which guarantees successful decoding in the worst case. Hence, the problem studied in this paper can be regarded as the probabilistic variant of the dual problem of the reconstruction problem. Yet another a highly related problem is the one of the \emph{trace reconstruction problem}~\cite{BKK04,DOS17,duda2016fundamental,HPP18,HMP08,NP17,PZ17}. The most relevant works to our study are the recent ones~\cite{srinivasavaradhan2018maximum,srinivasavaradhan2019symbolwise}, where decoding algorithms for maximum likelihood are presented for a fixed number of channels. 

\section{The Deletion and Insertion Channels}\label{sec:del_ch}
In this section we establish several basic results for the deletion channel with multiple instances. We start with several useful definitions. 
For two words $\bfx,\bfy\in\Sigma_q^*$, the number of times that $\bfy$ can be received as a subsequence of $\bfx$ is called the \emph{embedding number of $\bfy$ in $\bfx$} and is defined by
$$\emb(\bfx;\bfy) = |\{ I\subseteq [|\bfx|] \ | \ \bfx_I=\bfy\}|.$$
Note that if $\bfy$ is not a subsequence of $\bfx$ then $\emb(\bfx;\bfy)=0$. The embedding number has been studied in several previous works; see e.g.~\cite{DBLP:journals/corr/abs-1802-00703,elzinga2008algorithms} and in~\cite{srinivasavaradhan2018maximum} it was referred as the \emph{binomial coefficient}. In particular, this value can be computed with quadratic complexity~\cite{elzinga2008algorithms}.

While the calculation of the conditional probability $\pr_\ch\{  \bfy\textmd{ rec. }| \bfx \textmd{ trans.}\}$ is a rather simple task for many of the known channels, it is not straightforward for channels which introduce insertions and deletions. The following basic lemma is well known and was also stated in~\cite{srinivasavaradhan2018maximum}, however it is presented here for the completeness of the results in the paper and since it will be used in our derivations to follow.
\begin{claim}\label{cl:emb}
For all $(\bfx,\bfy)\in(\Sigma_q^*)^2$, it holds that \vspace{-1ex}
$$\pr_{\del(p)}\{  \bfy\textmd{ rec. }| \bfx \textmd{ trans.}\} = p^{|\bfx|-|\bfy|}\cdot \emb(\bfx;\bfy),\vspace{-1ex}$$
$$\pr_{\ins(p)}\{  \bfy\textmd{ rec. }| \bfx \textmd{ trans.}\} = \left(\frac{p}{q}\right)^{|\bfy|-|\bfx|}\cdot\emb(\bfy;\bfx).\vspace{-1ex}$$
\end{claim}
According to Claim~\ref{cl:emb}, it is  possible to explicitly characterize the ML decoder for the deletion and insertion channels as described also in~\cite{srinivasavaradhan2018maximum}.
\begin{claim}
Assume $\bfc\in\cC\subseteq (\Sigma_q)^n$ is the transmitted word and $\bfy \in (\Sigma_q)^{\leq n}$ is the output of the deletion channel $\del(p)$, then
$$\cD_{\ML}(\bfy) = \argmax_{\bfc\in\cC}\{ \emb(\bfc;\bfy)\}.$$
Similarly, for the insertion channel $\ins(p)$, for $\bfy \in (\Sigma_q)^{\geq n}$,
$$\cD_{\ML}(\bfy) = \argmax_{\bfc\in\cC}\{ \emb(\bfy;\bfc)\}.$$
\end{claim}
In case there is more than a single instance of the deletion channel, the following claim follows.
\begin{claim}
Assume $\bfc\in\cC\subseteq (\Sigma_q)^n$ is the transmitted word and $\bfy_1,\ldots,\bfy_t \in (\Sigma_q)^{\leq n}$ are the output words from $\del(p)$, then\vspace{-1ex}
$$\cD_{\ML}(\bfy_1,\ldots,\bfy_t) = \argmax_{\bfc\in\cC\cap \sups(\bfy_1,\ldots,\bfy_t)}\bigg\{\prod_{i=1}^t \emb(\bfc;\bfy_i)\bigg\},\vspace{-1ex}$$
and for the insertion channel $\ins(p)$, for  $\bfy_1,\ldots,\bfy_t \in (\Sigma_q)^{\geq n}$, \vspace{-1ex}
$$\cD_{\ML}(\bfy_1,\ldots,\bfy_t) = \argmax_{\bfc\in\cC\cap \subs(\bfy_1,\ldots,\bfy_t)}\bigg\{\prod_{i=1}^t \emb(\bfy_i;\bfc)\bigg\}.\vspace{-1ex}$$
\end{claim}
\begin{IEEEproof}
Every candidate to be considered in the ML decoder is a common supersequence of $\bfy_1,\ldots,\bfy_t$.  
Hence, $\cD_{\ML}(\bfy) = \bfc,$ where $\bfc\in\cC\cap \sups(\bfy_1,\ldots,\bfy_t)$ maximizes\vspace{-1.5ex}
$$\pr_{\del(p)}\{  \bfy_1,\ldots,\bfy_t \textmd{ rec. }| \bfc \textmd{ trans.}\} =  \prod_{i=1}^tp^{|\bfc|-|\bfy_i|}\cdot \emb(\bfc;\bfy_i).\vspace{-1.5ex}$$
Since $\prod_{i=1}^tp^{|\bfc|-|\bfy_i|}$ is the same for all candidates $\bfc$, the statement holds. A similar proof holds for the insertion channel. \vspace{-1ex}
\end{IEEEproof}

Note that since there is more a single channel, when the goal is to minimize the average decoding error probability, the ML decoder does not necessarily have to output a codeword but any word that minimizes the average decoding error probability. Thus, for the rest of the paper, when discussing the average decoding error probability it is assumed that the ML decoder can output \emph{any} word and not necessarily a codeword from $\cC$. Thus, we get the following claim. \vspace{-1.5ex}
\vspace{-.5ex}\begin{claim}\label{cl:ML}
Assume $\bfc\in\cC\subseteq (\Sigma_q)^n$ is the transmitted word and $\bfy_1,\ldots,\bfy_t \in (\Sigma_q)^{\leq n}$ are the output words from $\del(p)$, then\vspace{-1ex}
$$\cD_{\ML}(\bfy_1,\ldots,\bfy_t) = \argmax_{\bfx\in\sups(\bfy_1,\ldots,\bfy_t)}\bigg\{p^{|\bfx|\cdot t}\prod_{i=1}^t \emb(\bfx;\bfy_i)\bigg\},\vspace{-1ex}$$
and for the insertion channel $\ins(p)$, for  $\bfy_1,\ldots,\bfy_t \in (\Sigma_q)^{\geq n}$, \vspace{-1ex}
$$\cD_{\ML}(\bfy_1,\ldots,\bfy_t) = \argmax_{\bfx\in\subs(\bfy_1,\ldots,\bfy_t)}\bigg\{p^{|\bfx|\cdot t}\prod_{i=1}^t \emb(\bfy_i;\bfx)\bigg\}.\vspace{-1ex}$$
\end{claim}

Assume $\cC$ is $\Sigma_q^n$. The average decoding failure probability of the ML decoder over the deletion channel $\del(p)$ with $t$ instances is denoted by $\pfail(\del(p),t,\Sigma_q^n,\cD_{\ML})$ and shortly $\pfail(q,p,t)$. Similarly, the average decoding error probability is $\perr(\del(p),t,\Sigma_q^n,\cD_{\ML},d_L)$ and shortly $\perr(q,p,t)$. If $t=2$ it will be removed from the notations. 

Our main goal in the rest of the paper is to calculate close approximations for $\pfail(q,p,t)$ and $\perr(q,p,t)$ when $t=2$. Note that a lower bound on these probabilities is $p^t$ since if the same symbol is deleted in all of the channels, then it is not possible to recover its value and thus it will be deleted also in the output of the ML decoder. This was already observed in~\cite{srinivasavaradhan2018maximum} and in their simulation results. In the next section, we will analyze these probabilities for the special case of $t=2$, when the code is $\Sigma_q^n$, the VT code\cite{VT}, and the SVT code~\cite{7837631}.  

Complexity wise, it is well known that the time complexity to calculate the list of SCSs and LCSs is quadratic~\cite{itoga1981string}. This will also be the complexity to calculate the embedding numbers~\cite{elzinga2008algorithms} and thus the complexity of the ML decoder will be quadratic for $t=2$. The main idea behind these algorithms uses dynamic programming in order to calculate the list of SCSs, LCSs, and embedding numbers for every prefixes of the given words. However, when calculating for example the SCSs for $\bfy_1$ and $\bfy_2$ it is already known that $\mathsf{SCS}(\bfy_1,\bfy_2) \leq n$. Hence, it is not hard to observe that (see e.g.~\cite{apostolico1992fast}) many paths corresponding to prefixes which their length difference is greater than $d_1+d_2$ can be eliminated, when $d_1,d_2$ is the number of deletions in $\bfy_1,\bfy_2$, respectively. In particular, when $d_1$ and $d_2$ are fixed then the time complexity is linear. In our simulations we used this improvement when implementing the ML decoder. Other improvments and algorithms of the ML decoder are discussed in~\cite{srinivasavaradhan2018maximum,srinivasavaradhan2019symbolwise}.

\section{Two Deletion Channels}\label{sec:two deletions}

In this section we consider only the case of two deletion channels and prove in Theorem~\ref{th:2ch} an approximation for the average decoding error probability in the form of
$$\perr(q,p) \approx \frac{3q-1}{q-1}p^2 + O(p^3).$$
As mentioned in Section~\ref{sec:del_ch}, a lower bound on the value of $\perr(q,p,t)$ is $p^t$. This lower bound is indeed not tight since if symbols from the same run are deleted then the outputs of the two channels of this run are the same and it is impossible to recover that this run experienced a deletion in both of its copies. The probability of deletions due to runs is denoted by $\prun(q,p)$ and the next lemma approximates this probability. 
\vspace{-1ex} \begin{lemma} \label{lm:2ch_run_del}
For the deletion channel $\del(p)$, it holds that 
$$\prun(q,p)  \approx \frac{q+1}{q-1}p^2.$$
\end{lemma}\vspace{-1ex} 
\begin{IEEEproof} 
Given a run of length $r$, the probability that both of its copies have experienced a deletion is at roughly $(rp)^2$. Furthermore, the occurrence probability of a run of length exactly $r$ is $\left(\frac{1}{q}\right)^{r-1}\cdot\frac{q-1}{q}$. Thus, for $n$ large enough, the error probability is approximated by 
\begin{align*}
\sum_{r=1}^{\infty} (rp)^2  \left(\frac{1}{q}\right)^{r-1}&\cdot\frac{q-1}{q} = p^2\cdot \frac{q-1}{q} \sum_{r=1}^{\infty} r^2 \left(\frac{1}{q}\right)^{r-1} & \\
& = p^2\cdot \frac{q-1}{q} \frac{1+\frac{1}{q}}{(1-\frac{1}{q})^3} = p^2\cdot \frac{q(q+1)}{(q-1)^2}.& 
\end{align*}
The expected length of a run is given by 
$$\sum_{r=1}^{\infty} r \left(\frac{1}{q}\right)^{r-1}\cdot\frac{q-1}{q} =\frac{q-1}{q} \sum_{r=1}^{\infty} r \left(\frac{1}{q}\right)^{r-1} = \frac{q}{q-1}.$$
Hence, the expected number of runs in a vector of length $n$ is $n\cdot \frac{q-1}{q}$ and thus the approximated number of deletions due to runs is 
$$n\cdot \frac{q-1}{q} \cdot p^2\cdot \frac{q(q+1)}{(q-1)^2} = n p^2\cdot  \frac{q+1}{q-1},$$
which verifies the statement in the lemma. \end{IEEEproof}

However, runs are not the only source of errors in the output of the ML decoder. For example, assume the $i$-th, $(i+1)$-st symbols are deleted from the two channels. If the transmitted word $\bfx$ is of the form $\bfx = (x_1,\ldots,x_{i-1},0,1,x_{i+2},\ldots,x_n)$, then the two channels' outputs are $\bfy_1 = (x_1,\ldots,x_{i-1},0,x_{i+2},\ldots,x_n)$ and $\bfy_2 = (x_1,\ldots,x_{i-1},1,x_{i+2},\ldots,x_n)$. However, these two outputs could also be received upon deletions exactly in the same positions if the transmitted word is $\bfx' = (x_1,\ldots,x_{i-1},1,0,x_{i+2},\ldots,x_n)$. Hence, the ML decoder can output the correct word only in one of these two cases. Longer alternating sequences cause the same problem as well and the \emph{occurrence} probability of this event, denoted by $\palt(q,p)$, will be estimated in the next lemma. 
\vspace{-1ex} \begin{lemma} \label{lm:2ch_alt_del}
For the deletion channel $\del(p)$, it holds that 
$$\palt(q,p)  \approx 2p^2.$$
\end{lemma} \vspace{-1ex}
\begin{IEEEproof}
Assume there is a deletion in the first channel in the $i$-th position and the closest deletion in the second channel is $j>0$ positions apart, i.e., either in position $i-j$ or $i+j$. For simplicity assume it is in $(i+j)$-th position and $\bfx_{[i:j]}$ is an alternating sequence $ABAB\cdots$. Then, the 
same outputs from the two channels could be received if the transmitted word is the same as $\bfx$ but with changing the order of the alternating sequence, that is, the symbols of the word in the positions of $[i:j]$ are $BABA\cdots$. Therefore, the occurrence probability of this event can be approximated by
$$2p^2\cdot \sum_{j=1}^\infty \frac{q-1}{q}\cdot \frac{1}{q^{j-1}} = 2p^2,$$
where  $\frac{q-1}{q}\cdot \frac{1}{q^{j-1}}$ is the probability that $\bfx_{[i:j]}$ is any alternating sequence and the multiplication by 2 takes into account the cases of deletion in either position $i-j$ or $i+j$.
\end{IEEEproof}

At this point one may ask whether these two error events are the only dominant ones and indeed this question is answered in the affirmative, as stated in the following lemma.

\vspace{-1.5ex}
\begin{lemma} 
If there is a deletion in the first, second channel in position $i, i+j$, where $j>0$, respectively, and the sequence $\bfx_{[i:i+j]}$ is neither a run nor an alternating sequence, then these deletions are corrected successfully by the ML decoder. 
\end{lemma}

\vspace{-1.5ex}
We are now ready to show the following theorem on the Levenshtein error rate for the case of two channels. \vspace{-1.5ex}
\begin{theorem}\label{th:2ch}
The Levenshtein error rate for two deletion channels is approximated by 
\vspace{-1.5ex} $$\hspace{-0.3ex}\perr(q,\hspace{-0.25ex}p)\hspace{-0.35ex} \approx\hspace{-0.35ex} \prun(q,\hspace{-0.25ex}p) \hspace{-0.3ex}+\hspace{-0.3ex} \palt(q,\hspace{-0.25ex}p) \hspace{-0.3ex}+\hspace{-0.3ex}O(p^3) \hspace{-0.3ex}=\hspace{-0.3ex} \frac{3q\hspace{-0.3ex}-\hspace{-0.3ex}1}{q\hspace{-0.3ex}-\hspace{-0.3ex}1}p^2\hspace{-0.3ex}+\hspace{-0.3ex}O(p^3).$$
\end{theorem} \vspace{-1.5ex}
\begin{IEEEproof}
The proof follows from the above few lemmas. Note that each run error translates to an increase of the Levenshtein distance by one. On each occurrence of the alternating event the decoder chooses the correct subsequence with probability 0.5 and every error increases the Levenshtein distance by two since it translates to one insertion and one deletion. Lastly, the $O(p^3)$ expression compensates for all other less dominant error events which introduce more than two deletions at  that are close to each other at least in one of the channels.
\end{IEEEproof}

Using these observations, we are also able to approximate the average decoding failure probability. \vspace{-2ex}
\begin{theorem}\label{th:2chF}
The average decoding failure probability is \vspace{-1ex}
$$\pfail(q,p) \approx  e^{-\frac{3q-1}{q-1}p^2n}.$$
\end{theorem} \vspace{-1.5ex}
\begin{IEEEproof}
This is the probability that there was neither a deletion error because of the runs nor errors because of the alternating sequences. 
Hence, this probability becomes 
\begin{align*} \vspace{-1ex}
& \left( 1- \prun(q,p) \right)^n \cdot \left( 1- \palt(q,p) \right)^n & \\
& \approx  \left( 1- (\prun(q,p) +  \palt(q,p)) \right)^n =  \left( 1-\frac{3q-1}{q-1}p^2 \right)^n & \\
& \approx e^{-\frac{3q-1}{q-1}p^2n}.& \end{align*}  \\[-2.5\baselineskip] \end{IEEEproof} 

So far we have discussed only the case in which the code $\cC$ is the entire space. However, the most popular deletion-correcting code is theVT code~\cite{VT}. Recently, SVT, an extension of the VT code has been proposed in~\cite{7837631} for the correction of burst deletions, where the goal of this code construction was to correct a deletion error that its location is known up to some roughly $\log(n)$ consecutive locations. However, this construction has been recently used in~\cite{8294216} to build a code that is specifically targeted for the reconstruction of a word that is transmitted through two single-deletion channels. 
Due to the relevance of correcting deletion and alternating errors, the decoding failure probabilities of these two codes is investigated in this work. We abbreviate the notation of $\pfail(\del(p),2,VT_n,\cD_{\ML}), \pfail(\del(p),2,SVT_n,\cD_{\ML})$ by $\pfail(VT_n,q,p),\pfail(SVT_n,q,p)$, respectively. The following theorem summarizes these results. 
\begin{theorem}\label{th:2ch_VT}
The average decoding failure probabilities for the VT and SVT codes are given by
\begin{small}
\begin{align*}
\pfail(VT_n,q,p&)  \approx  \left( 1- \prun(q,p) \right)^n \cdot \left( 1- \palt(q,p) \right)^n  & \\
& + \left( 1- \prun(q,p) \right)^{n} \cdot n \palt(q,p) \left( 1- \palt(q,p) \right)^{n-1}, & \\
& + n \prun(q,p)\left( 1- \prun(q,p) \right)^{n-1} \cdot \left( 1- \palt(q,p) \right)^{n} & \\
\pfail(SVT_n,q,&p)  \approx  \left( 1- \prun(q,p) \right)^n \cdot \left( 1- \palt(q,p) \right)^n  & \\
& + \left( 1- \prun(q,p) \right)^{n} \cdot n \palt(q,p) \left( 1- \palt(q,p) \right)^{n-1}.& 
\end{align*}
\end{small}
\end{theorem}
\begin{IEEEproof}
The proof follows from the observation that the VT code is capable of decoding either a single run error or a single alternating error. On the other hand, the shifted VT code is capable of only correcting a single alternating error; see~\cite{8294216} for more details. 
\end{IEEEproof}

\begin{figure}[h!]
		\centering
		\subfigure{\includegraphics[width=85mm]{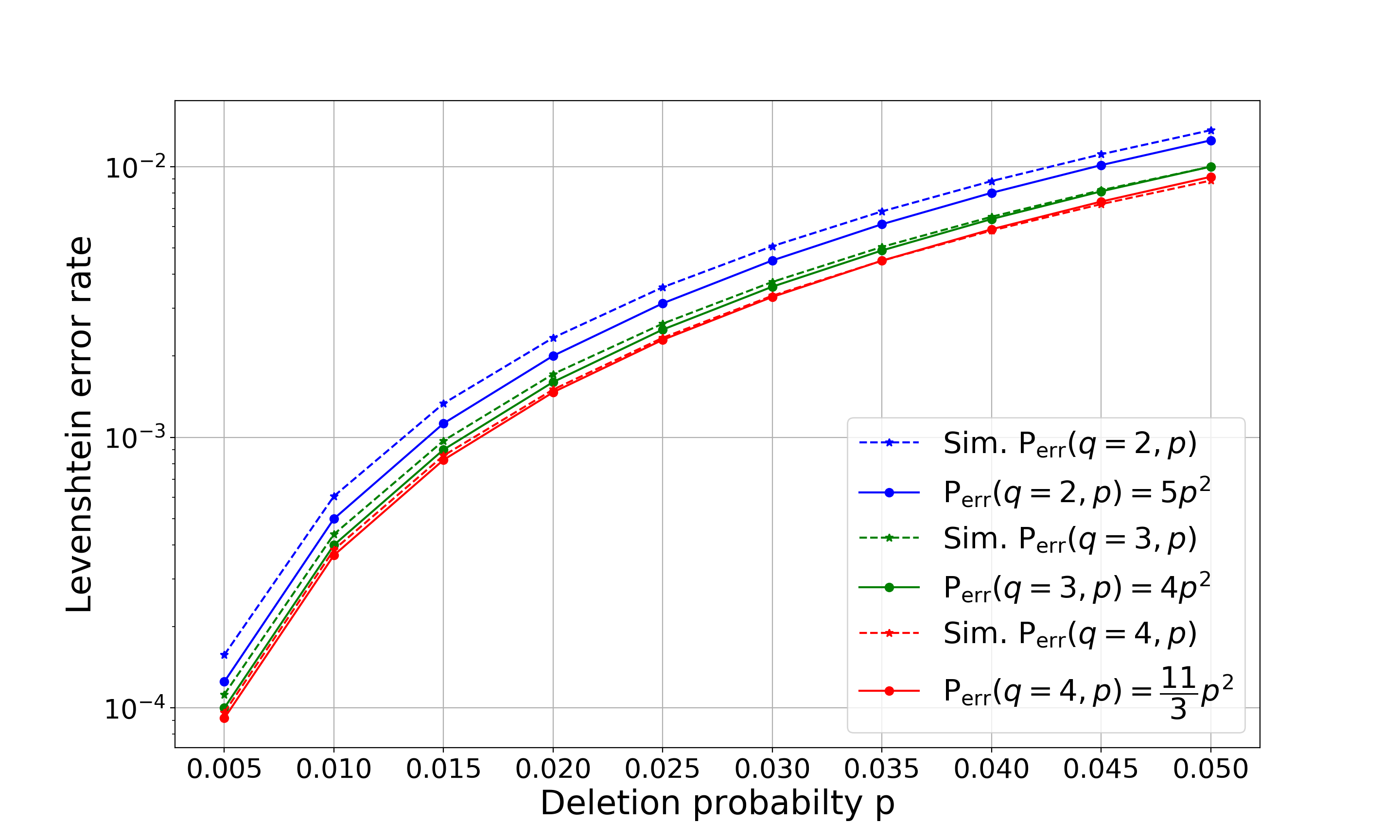}}\vspace{-2ex}
		\caption{Levenshtein error rate by the deletion probability $p$. These simulations verify Theorem~\ref{th:2ch}.} \label{fig:LER-Deletion}
	\end{figure}

\begin{figure}[h!] 
		\centering
		\subfigure{\includegraphics[width=85mm]{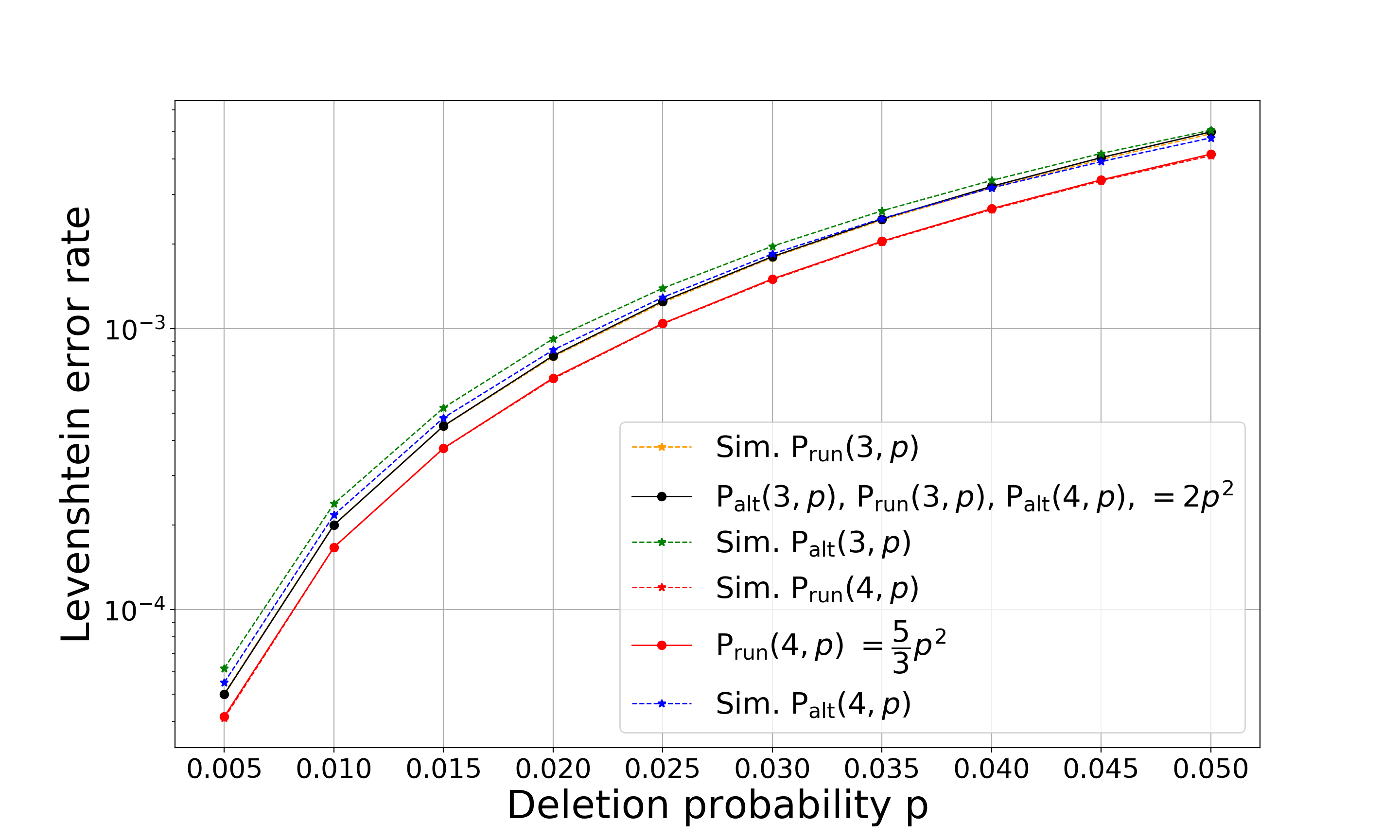}}\vspace{-2ex}
		\caption{Levenshtein error rate by the deletion probability $p$. These simulations verify Lemma~\ref{lm:2ch_run_del} and Lemma~\ref{lm:2ch_alt_del}.}\label{fig:runAltErrorsDel}
	\end{figure}

\begin{figure}[h!] 
		\centering
		\subfigure{\includegraphics[width=85mm]{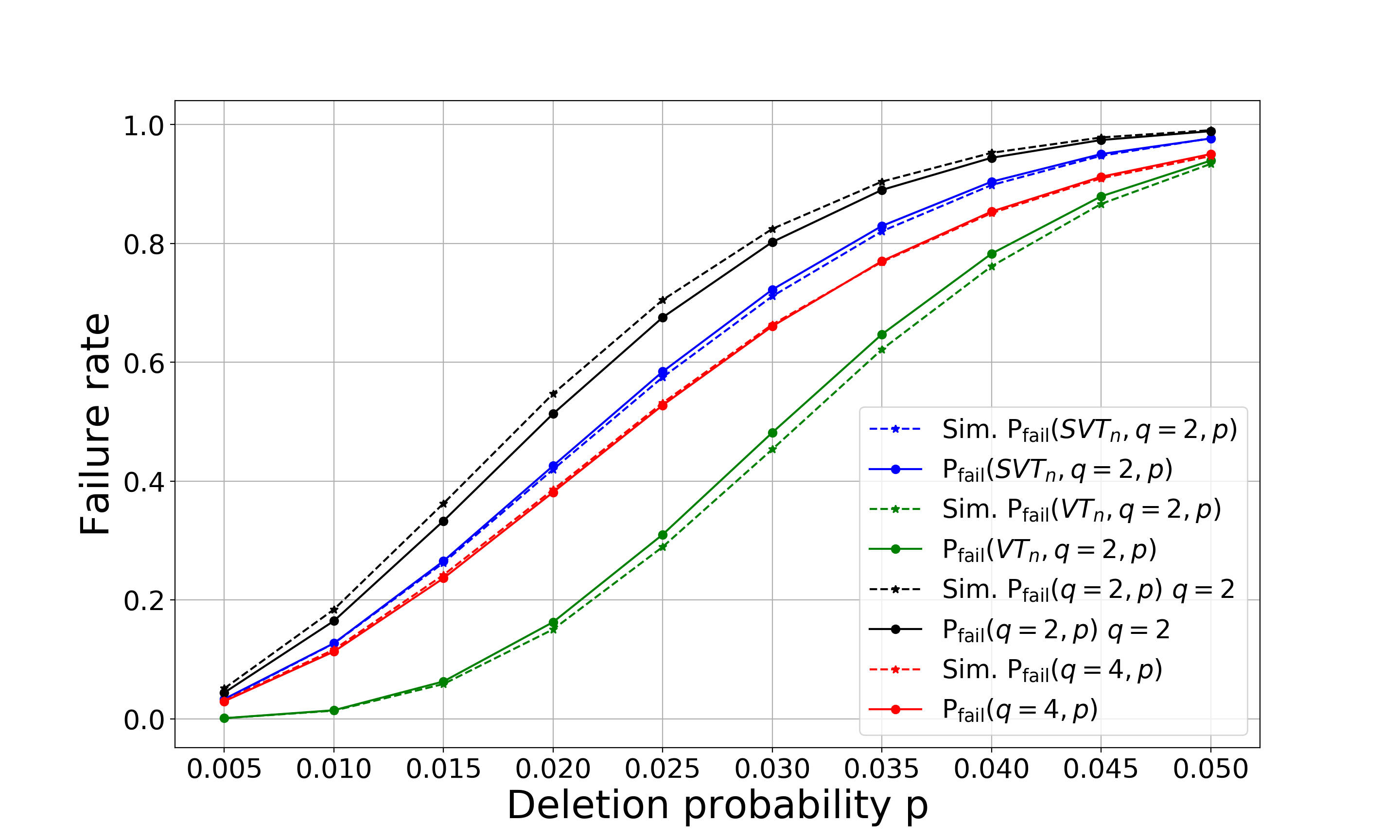}}\vspace{-2ex}
		\caption{Failure rate of the ML decoder, stratified by the coding scheme. These simulations verify Theorem~\ref{th:2chF} and Theorem~\ref{th:2ch_VT}.}\label{fig:failuerRate} \vspace{-3.75ex}
	\end{figure}

We verified the theoretical results presented in this section by the following simulations. These simulations were tested over words of length $n=450$ which were used to create two noisy copies given a fixed deletion probability $p\in[0.005,0.05]$. Then, the two copies were decoded by the ML decoder as described in Claim~\ref{cl:ML}. Finally, we calculated the Levenshtein error rate of the decoded word as well as the average decoding failure probability (referred here as \emph{failure rate}). Fig.~\ref{fig:LER-Deletion} plots the results of the Levenshtein error rate which confirms the probability expression $\perr(q,p)$ from Theorem~\ref{th:2ch}. Similarly, in Fig.~\ref{fig:runAltErrorsDel} we present separately the error probability $\palt(q,p), \prun(q,p)$, along with the corresponding value as calculated in Lemma~\ref{lm:2ch_run_del},~\ref{lm:2ch_alt_del}, respectively. Lastly, in Fig.~\ref{fig:failuerRate}, we simulated the ML decoder for the VT codes and the SVT codes and calculated the failure rate $\pfail(q,p)$. The implementation of the VT code was taken from~\cite{DBLP:journals/corr/abs-1906-07887}, and we modified this implementation for SVT codes. \vspace{-2ex}
	\vspace{-4ex}

\vspace{-2ex}
\section{Two Insertion Channels}\label{sec:ins}
This section continues the two-channel study but for for insertion case. In a similar manner to the deletion case, also here the dominant errors result from increasing the length of a run and error that results from the occurrence of an alternating sequence. We denote by $\perrin(q,p)$ the Levenshtein error rate of the ML decoder upon two instances of the insertion channel $\ins(p)$. Similarly, $\pfailin(q,p)$ is the average decoding failure probability and lastly 
$\prunin(q,p), \paltin(q,p)$ is the insertion probability due to runs, occurrence probability due to alternate sequences, respectively. The following theorem summarizes the results of this section. \vspace{-2ex}
\begin{theorem}\label{th:ins}
For the insertion channel $\ins(p)$, it holds that \vspace{-.5ex}
\begin{align*}
& \prunin(q,p) \approx \frac{q+1}{q(q-1)}p^2, \ \ \ \paltin(q,p)  \approx \frac{2}{q}p^2, \\
& \perrin(q,p) \approx \frac{3q-1}{q(q-1)}p^2 + O(p^3), \ \  \pfailin(q,p) \approx e^{-\frac{2}{q-1}p^2n}.
\end{align*}
\end{theorem}
Since the proof of this theorem repeats the same ideas as the ones for the deletion case we omit its details here. 

\begin{figure}[h!] 
		\centering
		\subfigure{\includegraphics[width=85mm]{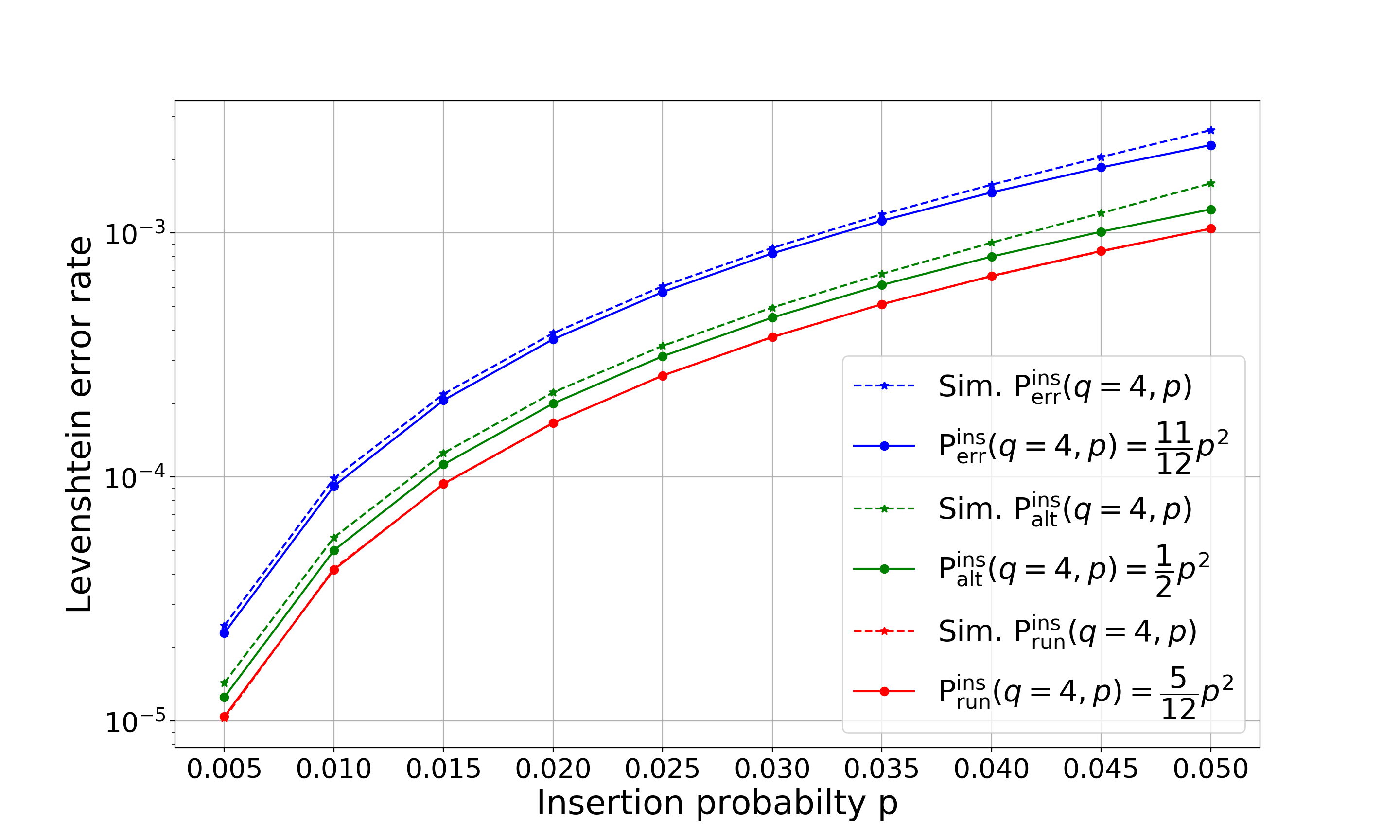}}\vspace{-2ex}
		\caption{Levenshtein error rate by the insertion probability $p$. This simulation verifies Theorem~\ref{th:ins}. }\label{fig:LER-Insertion}
	\end{figure}

The theoretical results of Theorem~\ref{th:ins} have also been verified by simulation results over words of length $n=500$, which were used to create two noisy copies with a given fixed insertion probability $p\in[0.005,0.05]$. Then, the two copies were decoded with the ML decoder according to Claim~\ref{cl:ML}. Lastly, we calculated and plotted in Fig.~\ref{fig:LER-Insertion} the Levenshtein error rate as well as the error rates from runs and alternating sequences.\vspace{-5ex}  

\vspace{-4ex}
\section{Conclustion}\label{sec:conc}
The main contribution of this paper is the study of the decoding error probability of the ML decoder for two deletion or insertion channels. While the results in the paper provide a significant contribution in the area of codes for insertions and deletions and sequence reconstruction, there are still several interesting problems
which are left open. Some of them are summarized as follows:
\begin{enumerate}
\item Study the non-identical channels case. For example two deletion channels with different probabilities $p_1$ and $p_2$.
\item Study the decoding error probability for more than two channels, both for insertions and deletions.
\item Study channels which introduce insertions, deletions, and substitutions.
\item Design coding schemes as well as complexity-efficient algorithms for the ML decoder in each case.
\end{enumerate}

\bibliographystyle{abbrv}
\bibliography{references}

\end{document}